\pdfoutput=1
\documentclass[sigconf,natbib=true,screen=True,authorversion=True]{acmart}
\AtBeginDocument{%
  \providecommand\BibTeX{{%
    \normalfont B\kern-0.5em{\scshape i\kern-0.25em b}\kern-0.8em\TeX}}}

\setcopyright{acmlicensed}
\copyrightyear{2024}
\acmYear{2024}
\acmDOI{10.1145/3626772.3657851}

\acmConference[SIGIR '24]{47th International ACM SIGIR
Conference on Research and Development in Information Retrieval}{July 14--18, 2024}{Washington D.C., USA}

\acmISBN{978-1-4503-XXXX-X/18/06}

\copyrightyear{2024}
\acmYear{2024}
\setcopyright{acmlicensed}\acmConference[SIGIR '24]{Proceedings of the 47th International ACM SIGIR Conference on Research and Development in Information Retrieval}{July 14--18, 2024}{Washington, DC, USA}
\acmBooktitle{Proceedings of the 47th International ACM SIGIR Conference on Research and Development in Information Retrieval (SIGIR '24), July 14--18, 2024, Washington, DC, USA}
\acmDOI{10.1145/3626772.3657851}
\acmISBN{979-8-4007-0431-4/24/07}

\usepackage{xcolor}
\usepackage{multirow}
\usepackage{array}
\usepackage{caption}
\usepackage{listings}
\usepackage{xspace}
\usepackage[most]{tcolorbox}
\usepackage{ifthen}
\usepackage{subcaption}
\usepackage{makecell}

\usepackage{tablefootnote}
\definecolor{requestIdcolor}{HTML}{F0B658}
\definecolor{querycolor}{HTML}{68C1DC}
\definecolor{titlecolor}{HTML}{85A1DD}
\definecolor{urlcolor}{HTML}{8D8171}
\definecolor{rankcolor}{HTML}{BD87CC}
\definecolor{yescolor}{HTML}{83D0A9}
\definecolor{nocolor}{HTML}{C36384}
\newtcolorbox{pill}[4][]{%
  enhanced,boxsep=1pt,top=1pt,left=1.5pt,bottom=0pt,right=2pt,arc=3pt,boxrule=1pt,leftrule=#4,#1
  colback=white,colframe=#3color,
  overlay unbroken and first ={\node[anchor=west,font=\sc\bfseries,xshift=-2pt,white] at (frame.west) {#2};}
}
\newcommand{\pillrow}[5]{%
  \begin{pill}{title}{title}{3em}\vphantom{Xg}#4\end{pill}&%
  \begin{pill}{url}{url}{1.95em}\vphantom{Xg}\dots\end{pill}&%
  \begin{pill}{rank}{rank}{2.8em}\vphantom{Xg}#1\end{pill}&%
  \ifthenelse{#2>0}{\begin{pill}{clicks}{yes}}{\begin{pill}{clicks}{no}}{3.4em}\vphantom{Xg}#2\end{pill}&%
  \ifthenelse{{\not\equal{#3}{N/A}}\and#3>0}{\begin{pill}{dwelltime}{yes}}{\begin{pill}{dwelltime}{no}}{5.5em}\vphantom{Xg}#3\end{pill}%
  \\[-10pt]%
  \multicolumn{5}{l}{\begin{pill}{bte}{title}{3em}\vphantom{X}#5\end{pill}\kern-10pt}\\%
}
\NewDocumentCommand{\anote}{}{\makebox[0pt][l]{$^*$}}

\newenvironment{citemize}{\begin{list}{$\bullet$}{\topsep=.2\smallskipamount\itemsep=0pt\parsep=0pt\labelwidth=.5em}}{\end{list}}

\usepackage[absolute]{textpos}
\begin{document}
\begin{textblock}{16}(0,0.1)\centerline{This paper was published at \textbf{SIGIR 2024} -- please cite the published version {\small\url{https://doi.org/10.1145/3626772.3657851}} instead.}\end{textblock}

\def\clickset{CWRCzech\xspace}
\def\tablabel{Table}
\def\figlabel{Figure}

\hyphenation{CWRCzech}

\title{\clickset: 100M Query-Document Czech Click Dataset and Its Application to Web Relevance Ranking}

\author{Josef Vonášek}
\orcid{0009-0006-9429-9278}
\affiliation{
  \institution{Seznam.cz}
  \streetaddress{Radlická 3294/10, Smíchov}
  \city{Prague}
  \country{Czech Republic}
  \postcode{15000}}
\email{josef.vonasek@firma.seznam.cz}

\author{Milan Straka}
\orcid{0000-0003-3295-5576}
\affiliation{
  \institution{Charles University, Faculty of Mathematics and Physics}
  \streetaddress{Malostranské náměstí 25}
  \city{Prague}
  \country{Czech Republic}
  \postcode{11800}}
\email{straka@ufal.mff.cuni.cz}

\author{Rostislav Krč}
\orcid{0000-0001-6772-2575}
\affiliation{
  \institution{Seznam.cz}
  \streetaddress{Radlická 3294/10, Smíchov}
  \city{Prague}
  \country{Czech Republic}
  \postcode{15000}}
\email{rostislav.krc@firma.seznam.cz}

\author{Lenka Lasoňová}
\orcid{0009-0003-7524-0367}
\affiliation{
  \institution{Seznam.cz}
  \streetaddress{Radlická 3294/10, Smíchov}
  \city{Prague}
  \country{Czech Republic}
  \postcode{15000}}
\email{lenka.lasonova@firma.seznam.cz}

\author{Ekaterina Egorova}
\orcid{0000-0001-5398-4371}
\affiliation{
  \institution{Seznam.cz}
  \streetaddress{Radlická 3294/10, Smíchov}
  \city{Prague}
  \country{Czech Republic}
  \postcode{15000}}
\email{ekaterina.egorova@firma.seznam.cz}

\author{Jana Straková}
\orcid{0000-0003-0075-2408}
\affiliation{
  \institution{Charles University, Faculty of Mathematics and Physics}
  \streetaddress{Malostranské náměstí 25}
  \city{Prague}
  \country{Czech Republic}
  \postcode{11800}}
\email{strakova@ufal.mff.cuni.cz}

\author{Jakub Náplava}
\orcid{0000-0003-2259-1377}
\affiliation{
  \institution{Seznam.cz}
  \streetaddress{Radlická 3294/10, Smíchov}
  \city{Prague}
  \country{Czech Republic}
  \postcode{15000}}
\email{jakub.naplava@firma.seznam.cz}

\renewcommand{\shortauthors}{Josef Vonášek et al.}

\begin{abstract}
  We present \clickset, \textbf{C}lick \textbf{W}eb \textbf{R}anking dataset for \textbf{Czech}, a 100M query-document Czech click dataset for relevance ranking with user behavior data collected from search engine logs of Seznam.cz. To the best of our knowledge, \clickset is the largest click dataset with raw text published so far. It provides document positions in the search results as well as information about user behavior: 27.6M clicked documents and 10.8M dwell times. In addition, we also publish a manually annotated Czech test for the relevance task, containing nearly 50k query-document pairs, each annotated by at least 2 annotators. Finally, we analyze how the user behavior data improve relevance ranking and show that models trained on data automatically harnessed at sufficient scale can surpass the performance of models trained on human annotated data. \clickset is published under an academic non-commercial license and is available to the research community at \url{https://github.com/seznam/CWRCzech}.
\end{abstract}



\maketitle

\section{Introduction}

In the field of information retrieval (IR), the task of relevance ranking is to determine the degree of relevance of documents or items with respect to a particular query. In order to accommodate lengthier, more naturally phrased queries as opposed to keywords, modern relevance has moved away from rule-based approaches 
towards pre-trained language models~\cite{Nogueira2019MultiStageDR,nogueira-etal-2020-document,MacAvaney2019,Dai2019,Li2023PARADE,Yilmaz2019,SZN_SmallECzech}. However, the effectiveness of these models depends heavily on the availability of extensive training data.

Although human relevance annotation provides high-quality training data, it is costly and time-consuming. Harnessing user behavior data collected in production offers a robust, cost-effective option; nevertheless, such query-document click datasets are not routinely published or available for academic non-commercial use at scale, much less so in non-English languages. To date, a few large-scale datasets containing user behavior data in the search domain have been released~\cite{Craswell2020Orcas,Rekabsaz2021TripClick,Serdyukov2014Yandex,Zheng2018SogouQCL,Zou2022Baidu}.%

In order to contribute to the research area of user behavior in the context of relevance ranking, we publish \clickset (\textbf{C}lick \textbf{W}eb \textbf{R}anking dataset for \textbf{Czech}), a new dataset of 100M query-document pairs in the Czech language derived from search engine logs of Seznam.cz.\footnote{\url{https://search.seznam.cz/}} It contains not only positive examples but also negative ones (offered but not clicked), which makes it a valuable resource for model training. To our knowledge, the presented dataset is by far the largest click dataset with raw text. 

To provide a proper evaluation benchmark, we have also manually annotated and released a representative Czech test set for the relevance task, containing circa 50k query-document pairs, each manually annotated by at least 2 annotators to ensure high quality of the annotations.

To showcase the research potential of the corpus for the relevance ranking research, we analyze and experimentally validate to what degree such automatically collected, inherently noisy user behavior data contribute to training large language models for the relevance ranking task in comparison with human-annotated data. We find that user data generated automatically at sufficient scale challenge the performance of human annotations when evaluated on in-domain relevance ranking.
Our contributions are:
\begin{itemize}
    \item a new, large query-document click dataset \clickset for Czech relevance ranking containing 100M query-document pairs of user behavior collected in production,
    \item manually annotated Czech test set of around 50k query-document pairs, each annotated by at least 2 annotators,
    \item model analysis and experimental validation of the contribution of automatically harnessed query-document click data for relevance ranking.
\end{itemize}

\clickset is published under a non-commercial license and is available at
\textbf{\url{https://github.com/seznam/CWRCzech}}.

\section{Related Work}
\label{sec:related_work}

\subsubsection*{Datasets}

Two primary methods are used to create relevance datasets. The first involves human annotators who manually assess the relevance of each document. The second method directly utilizes user behavior data collected during the use of the company's services.

For instance, Microsoft's MS MARCO dataset~\cite{Nguyen2016MSMarco} includes 8.8 million query-document pairs with human-provided relevance annotations. There are also non-English relevance datasets like the Chinese T2Ranking~\cite{Xie2023T2Ranking} consisting of 2 million annotated pairs, and the Czech DaReCzech, featuring 1.6 million pairs.

The large-scale English click datasets are often compiled directly from search engine logs. Notable examples are the AOL~\cite{Pass2006Picture} and the  MSN~\cite{Zhang2006Some} datasets with millions of queries. Microsoft recently released the ORCAS dataset~\cite{Craswell2020Orcas} containing 18.8 million query-document pairs. As an example of a domain-specific dataset, one can look at TripClick with 1.3 million pairs~\cite{Rekabsaz2021TripClick}, acquired from a health web search engine. The most notable non-English datasets provided by other search engine companies include Russian Yandex-WSCD~\cite{Serdyukov2014Yandex} with 35 million search sessions and anonymized queries, or Chinese Sougou-QCL~\cite{Zheng2018SogouQCL} and Baidu-ULTR~\cite{Zou2022Baidu} datasets with 12.2 million pairs and 1.2 billion pairs respectively, with queries and documents anonymized using a proprietary dictionary.

\looseness-1
The type of information provided in a click dataset is vital and holds greater importance than size alone when training neural models for retrieval and ranking. A detailed comparison of different click datasets to \clickset is discussed in Section~\ref{sec:click_data}. Notably, click datasets differ in the level of query token anonymization, ranging from dataset replacing all words with randomized IDs~\cite{Serdyukov2014Yandex}, through partial word replacement~\cite{Zou2022Baidu}, to datasets with original, raw text, as is the case of ORCAS~\cite{Craswell2020Orcas} and our dataset. Proper handling of tail, sensitive or harmful queries is important, and we cover this in Section~\ref{sec:data}.

\goodbreak

\subsubsection*{Relevance Ranking Approaches}

Traditional non-neural techniques based on term frequency, such as BM25~\cite{robertson-1994-okapi} or term frequency-inverse document frequency (TF-IDF) representations, were recently widely replaced or extended by neural methods based on the Transformer architecture~\cite{NIPS2017attention}, such as BERT~\cite{devlin-etal-2019-bert}. Pretrained language models have demonstrated their capability for dense retrieval~\cite{Zhao2022DensePLM} in application domains such as E-commerce~\cite{Yao2021Alibaba}, recommendation~\cite{Xie2023Reweighting}, online advertising~\cite{Yang2021HybridAds}, or web search~\cite{SZN_SomeLikeItSmall}.

Unlabeled data for contrastive learning in web search originate from logged user interactions with the search engine such as clicks or time spent on a particular result (dwell time). These feedback signals can be viewed as an approximation of relevance and used as positive labels. To increase training effectiveness, the larger unlabeled log data can be combined with smaller human-annotated relevance sets for further fine-tuning. 

Click data are, for example, frequently coupled with annotated relevance to train rankers~\cite{Yao2021Alibaba, Lin2021Modeling}. However, clicks also suffer from position bias when items in the top positions receive more exposure and therefore have higher click probability than bottom items~\cite{Chuklin2022ClickModels}. Position bias is commonly considered in models for click-through rate prediction, but it can also improve the quality of training data for relevance models~\cite{Yao2021Alibaba}.

Another option is to incorporate user's dwell time~\cite{Yi2014Beyond, Kim2014Modeling} into the training objective. Short interaction with the clicked result may indicate poor relevance and vice versa. Improved performance when high-quality items are placed in the top positions is observed when clicks are reweighted by the normalized dwell time~\cite{Xie2023Reweighting}.

\begin{table*}[t]
\centering
\small
  \caption{Visualization of a query from the \clickset dataset. The query and the documents have been
  translated to English for better understanding, and only excerpts of the body text extracts are shown.}
  \vspace{-5pt}
\setlength{\tabcolsep}{1pt}\begin{tabular}{p{11.76cm}p{1.03cm}p{1.15cm}p{1.3cm}p{2.25cm}}
\toprule
  \multicolumn{5}{l}{%
    \begin{pill}[hbox,]{requestId}{requestId}{5.3em}\vphantom{Xg}\texttt{18242939}\end{pill}\kern2pt
    \begin{pill}[hbox,]{\vphantom{X}query}{query}{3.3em}\vphantom{Xg}automatic parking\end{pill}}\\[-2pt]
\midrule
  \pillrow{0}{1}{116}{Automatic parking is not just a privilege of luxury cars - roadblog.cz}{Arrive at a parking spot, press a button, and let the car park itself. Such a feature is now available in accessible cars as well. You might say, that \dots\kern-1pt}[5pt]
  \pillrow{1}{0}{0}{Drivers do not use automatic parking, even though it is better than a human – AutoRevue.cz}{Automatic parking also uses 47 \% fewer maneuvers and corrections, and there was not a single instance of contact with another vehicle, unlike \dots}[5pt]
  \pillrow{3}{0}{0}{Automatic parking - cars that park themselves | OneTwoGo Car Rental}{Parallel parking is a struggle for many drivers, especially in big cities. Given that parking space is significantly limited by cars on crowded streets \dots\kern-2pt}[5pt]
  \pillrow{5}{0}{0}{Automatic Parking - Glossary of Terms - Electric Cars | Alza.cz }{Most automobile manufacturers provide the feature of automatic parking as an additional equipment option, even for their more affordable \dots}[5pt]
  \pillrow{7}{1}{N/A}{Description and principle of operation of the automatic parking system - AvtoTachki}{Parking a car is perhaps the most common maneuver that causes difficulties for drivers, especially inexperienced ones. But it was not so long \dots}
\bottomrule
\end{tabular}
\label{tab:data_example}
\end{table*}

\begin{table*}[t]
  \centering
  \caption{Comparison of \clickset to other publicly available click datasets for ranking. The table displays the number of unique queries, documents, total query-document (Q-D) pairs, and the search results sessions. Information about the data contained in each dataset and their languages is provided. $^*$TripClick comes from healthcare domain contrary to other listed web search datasets. $^\S$The tokens are represented as identifiers to a private dictionary; therefore, they cannot be used with pre-trained language models. $^\dagger$Dataset contains click-model generated relevance labels. $^\ddagger$ Dataset contains click sequence, displayed time/count, and others.}
  \vspace{-5pt}
  \label{tab:datasets}
  \setlength{\tabcolsep}{3pt}
  \begin{tabular}{p{2.8cm}rrrcccccccc}%
  \toprule

    \textbf{Dataset} & \makecell[l]{\textbf{Q-D}\\\textbf{pairs}} &\textbf{Queries} & \textbf{Docs} & \textbf{Language} & \makecell{\textbf{Query}\\\textbf{text}} &  \makecell{\textbf{Doc}\\\textbf{title}} & \makecell{\textbf{Doc}\\\textbf{body}} & \textbf{Clicks} & 
    \makecell{\textbf{Dwell}\\\textbf{time}} & \textbf{Rank} & \makecell{\textbf{Additional}\\\textbf{Information}} \\

  \midrule

    ORCAS~\cite{Craswell2020Orcas} &  18.8M &10.4M & 1.4M & English & raw &  raw &raw & \checkmark & - & - & - \\ %

    TripClick~\cite{Rekabsaz2021TripClick} &  5.3M &1.6M & 2.3M & English\anote & raw &  raw &raw & \checkmark & - & \checkmark & - \\ %

    TianGong-ULTR~\cite{Ai2018TiangongULTR1, Ai2018TiangongULTR2} &  &3.4K & 333.8K & Chinese & raw &  raw &raw & \checkmark & - & \checkmark & - \\ %
    
    Sougou-QCL~\cite{Zheng2018SogouQCL} & 12.2M &0.5M & 9.0M & Chinese & raw &  raw &raw & - & - & - & \checkmark\rlap{$^\dagger$} \\ %

    Baidu-ULTR~\cite{Zou2022Baidu} & & 383.4M & 1.3B & Chinese & private$^\S$ &  private$^\S$ &private$^\S$ & \checkmark & \checkmark & \checkmark & \checkmark\rlap{$^\ddagger$} \\ %

    Yandex-WSCD~\cite{Serdyukov2014Yandex} & 667.2M & 21.1M & 70.3M & Russian & private$^\S$ &  - & - & \checkmark & - & \checkmark & - \\ %
    
  \midrule

    \clickset & 100.0M & 2.7M & 8.4M & Czech & raw &  raw&$\le$230 chrs & \checkmark & \checkmark & \checkmark & - \\

  \bottomrule

\end{tabular}
\end{table*}

\section{Datasets}
\label{sec:data}

\subsection{\clickset Click Dataset}
\label{sec:click_data}

\clickset (\textbf{C}lick \textbf{W}eb \textbf{R}anking dataset for \textbf{Czech}) is a new Czech click dataset comprising 100 million query-document pairs derived from search engine logs of Seznam.cz. It contains over 2.7 million distinct queries and over 8.4 million documents. The queries come from requests collected over an extended period of time. We only selected the queries in Czech (according to the internal classifier) that were identified by the search engine as an informational intent~\cite{taxonomy_web_search}. Informational intent queries seek to acquire information (e.g., ``how to boil an egg'') and tend to be more naturally phrased, as opposed to navigational or transactional intent queries. \tablabel~\ref{tab:data_example} provides an example of a search results record for a user query. %

When constructing the dataset, our goal was to avoid sensitive information, user identification, and harmful content. To this end, we adhered to the following protocol: All queries classified as porn or obscene were filtered; as well as bot queries. To prevent an accidental leak of numerical information, such as credit card numbers, only queries with alphabetical characters were selected. 
Sessions were not merged by user ids for privacy reasons.
Finally, each query had a minimal occurrence in 5 unique requests within the specified time frame (i.e., the same query was requested by at least 5 users) to ensure anonymization and to prevent potential identification of specific users or their sensitive information.

In order to enhance query variability, the maximal number of unique requests for each query was limited to 15 (i.e., if the same query was requested by more than 15 users, we choose 15 requests at random), yielding 22.1M unique requests over the entire dataset. For inclusion in the dataset, each request was associated with documents extending up to the last click or up to the fifth position, whichever was greater. To concentrate on more complex inquiries, the dataset was curated to include only queries comprising a minimum of 10 characters.

The dataset contains the following columns:
\begin{citemize}
  \item \textit{requestId}: Id of the particular request with a single query.%
  \item \textit{query}: User query with corrected typos and added diacritical marks.
  \item \textit{url}: Document URL. 
  \item \textit{title}: Words from the document classified by the search engine as a title.
  \item \textit{bte}: Body text extract, i.e., document body snippet processed by the internal search engine model and trimmed to 230 characters (snippet size complying with fair use). It is empty for the webpages that block search engines or prohibit usage of their contents for GPT training.
  \item \textit{rank}: Position of the document in the search results page. Indices may be absent in cases where a document was no longer indexed at the time of dataset creation.
  \item \textit{clicks}: The number of clicks on a given document in given search results. 
  \item \textit{dwellTime}: Time in seconds spent in the clicked document page before the user returned to the search results page. This information is not always available, typically for the last click in the search results.
\end{citemize}

\begin{figure}[t]
  \centering
  \includegraphics[width=.75\hsize]{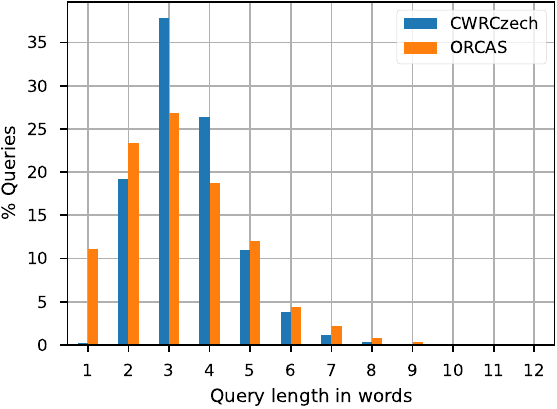}
  \vspace{-10pt}
  \caption{Comparison of query length in words (separated by a blank space) between \clickset and ORCAS~\cite{Craswell2020Orcas}.}
  \label{fig:dataset_stats}
  \Description{Query word count.}
  \vspace{5pt}
\end{figure}

\tablabel~\ref{tab:datasets} presents statistics summarizing the new \clickset dataset in comparison with other click datasets. Among datasets with readable text, i.e., without identifiers into a private dictionary, \clickset is the largest one, thus it is significantly larger than ORCAS \cite{Craswell2020Orcas}. The total number of documents is higher, however, it has a lower count of unique queries. 
The median query length in \clickset is 3 words and the average length is 3.48 words. Contrary to ORCAS, long queries and one-word queries are less common in \clickset. The distribution is illustrated in \figlabel~\ref{fig:dataset_stats}. Moreover, unlike ORCAS, \clickset contains whole search results with both clicked and unclicked documents as well as information about document rank and user dwell time. This provides a detailed insight into user interaction with search results that can be utilized for more precise relevance estimates and efficient ranker training.

\begin{figure}[t]
    \begin{subfigure}{\hsize}
      \centering
      \includegraphics[width=.612\hsize]{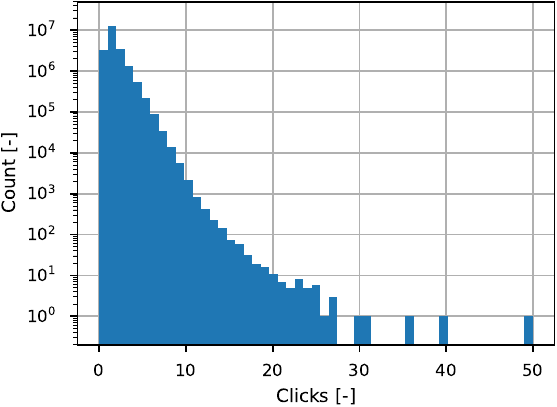}
      \vspace{-5pt}
      \caption{Number of clicks per requestId.}
      \vspace{5pt}
    \end{subfigure}
    \begin{subfigure}{\hsize}
      \centering
      \includegraphics[width=.646\hsize]{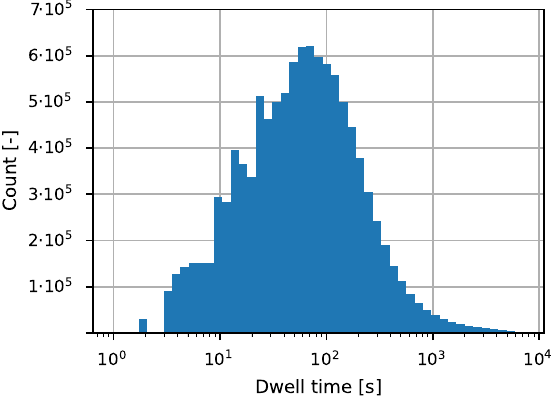}
      \vspace{-5pt}
      \caption{Distribution of dwell times for clicked documents.}
      \vspace{5pt}
    \end{subfigure}
    \begin{subfigure}{\hsize}
      \centering
      \includegraphics[width=.6715\hsize]{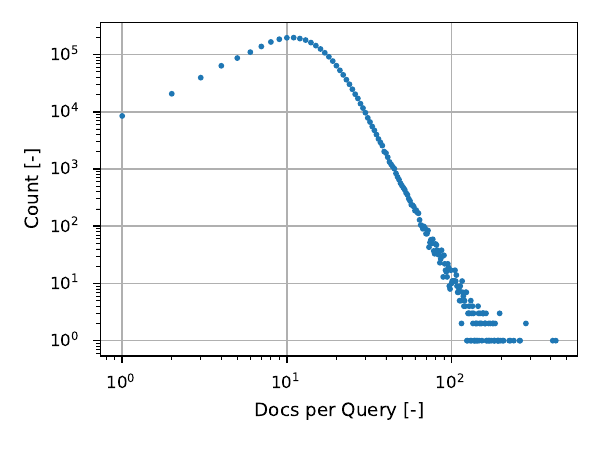}
      \vspace{-5pt}
      \caption{Number of documents per query.}
      \vspace{5pt}
    \end{subfigure}
    \begin{subfigure}{\hsize}
      \centering
      \includegraphics[width=.629\hsize]{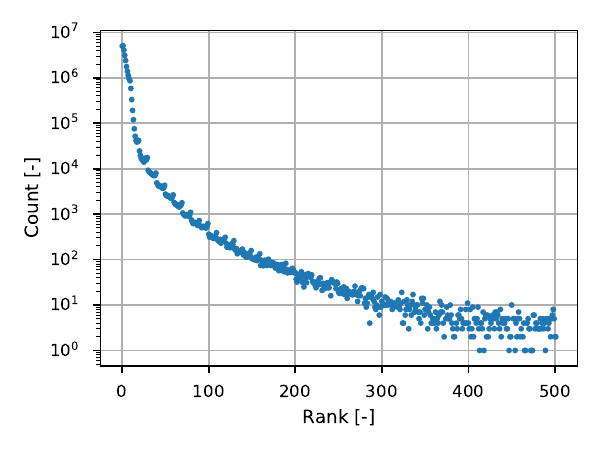}
      \vspace{-5pt}
      \caption{Distribution of ranks for clicked documents.}    
      \vspace{-2pt}
    \end{subfigure}
    \caption{Statistics of clicks, dwell times, and ranks within the \clickset dataset.}
    \label{fig:dataset_dist}
\end{figure}

\figlabel~\ref{fig:dataset_dist} illustrates some of the \clickset statistics: the distribution of the number of clicks per query-document pair, the distribution of dwell times where they are available, the number of documents per query, and the correlation between the rank of the document and the number of clicks it receives. %
Out of all individual query-document pairs, 27.6M are clicked; after aggregation, 60\% of the query-document pairs received no clicks, 24\% were clicked exactly once, and 16\% of the documents received more than one click (see \figlabel~\ref{fig:dataset_dist}.a). Both clicks and non-clicks provide valuable information about user interaction with the search results. Clicks are commonly treated as a strong relevance signal, but non-clicks can be used, for example, for the construction of soft negative pairs (see Section~\ref{sec:soft_negatives_methods}).
User post-click behavior is equally significant. An example of such behavior is dwell time, which is provided in \clickset explicitly. Note that it is available for 10.8\% pairs with the mean value of 132.5 seconds and the median of 58 seconds (see \figlabel~\ref{fig:dataset_dist}.b). The rank of the document on a page provides an insight into a potential position bias. Number of documents per query peaks at 10 (\figlabel~\ref{fig:dataset_dist}.c) which is the size of the first results page. The probability of a click generally decreases with increasing rank and more than half of the clicks occur in the top three results (see \figlabel~\ref{fig:dataset_dist}.d). Results paging causes visible steps in the graph for ranks divisible by the search results length since users are more likely to click on top documents on each page.


As all European-based companies are required to comply with the General Data Protection Regulation (GDPR), most relevantly Act. 17 Right to Erasure (`Right to be Forgotten'), the corpus is available under a non-commercial license upon request to keep record of the corpus users for broadcasting the potential erasure requirements.

\subsection{\clickset Annotated Test Set}
\label{sec:annotated_test_set}

Along with the \clickset click dataset, we also publish a manually annotated Czech test set for model evaluation.

We retrieved queries from the 2023 search logs, and performed random selection, deduplication, filtering of only informational intent as in Section~\ref{sec:click_data}. Each query was manually scrutinized for safety, sensibility, and anonymity. This process resulted in 995 queries that were designated as the test set. Each query is paired with documents of varying relevance, ranging from highly relevant documents in the search results to those with little to no relevance, often found in the early stages of retrieval. The texts of the documents were trimmed to 230 characters to comply with fair use. In total, the test set comprises 49,945 rows of query-document pairs. On average, each query is linked to 50.20 documents, with a minimum of 31 and a maximum of 89, and 19.27\% of these documents are deemed relevant (\( label > 0.5 \), see Label Design in Section~\ref{sec:label}). There is no overlap between the annotated test set and the \clickset click set.

Each query-document pair underwent annotation by at least 2 annotators,\footnote{Our annotators were in-house expert employees, native speakers of Czech, and predominantly women. They were compensated based on the number of annotations they made and their pay was above the mean salary valid in 2023 in the relevant country.} who had the option to assign the pair one of the following four values indicating the usefulness of the document with respect to the query: ``useful'' (labeled as 1), ``slightly useful'' (0.66), ``almost useless'' (0.33), and ``useless'' (0). In case when two annotations considerably differed, i.e., one was ``(slightly) useful'' and the other ``(slightly) useless'', a third annotator was asked to provide another annotation. This happened in 10\% of the cases. The ultimate label for each pair was determined by calculating the median of the two or three assigned values.

\begin{table}[]
    \caption{Annotated test sets data summary. The table displays the total number query-document (Q-D) pairs, the number of unique queries, the average number of documents per query (Avg. D/Q), and the number of annotators per query-document pair (Num. Annots).}
    \label{tab:testset_data}
    \setlength{\tabcolsep}{4.7pt}
    \begin{tabular}{lccccccc}
        \toprule
        \textbf{Test Set Name} & \makecell[c]{\textbf{Q-D}\\\textbf{pairs}} & \makecell[c]{\textbf{Unique}\\\textbf{Queries}} & \makecell[c]{\textbf{Avg.}\\\textbf{D/Q}} & \makecell[c]{\textbf{Num.}\\\textbf{\llap{A}nnot\rlap{s}}} \\ %
        \midrule
        DaReCzech (test) & 64,466 & 2,323 & 27.75 & 1 \\
        \textit{- non-informational intents} & 54,899 & 1,609 & 34.12 & 1\\
        DaReCzech (dev) & 41,220 & \hphantom{0,}793 & 51.98 & 1 \\
        \textit{- informational intent} & \hphantom{0}4,828 & \hphantom{0,0}99 & 48.77 & 1\\
        \midrule
        CWRCzech & 49,945 & \hphantom{0,}995 & 50.20 & 2-3 \\
        \bottomrule
    \end{tabular}
\end{table}

\subsection{DaReCzech Dataset}
\label{sec:dareczech_test_set}

In addition to the aforementioned test set, our experiments also make use of a previously released Czech dataset DaReCzech~\cite{SZN_SmallECzech}. Like the \clickset test set, the DaReCzech test set is also human annotated. However, its annotation reliability is lower than that of \clickset as the labels are based on a single annotation for each query-document pair. Another key distinction between the two datasets is that DaReCzech encompasses a full range of user intents, not just informational ones. As shown in \tablabel~\ref{tab:testset_data}, there is a significant domain shift in intents between \clickset and DaReCzech test set, as circa 70\% of the DaReCzech test is non-informational intent. Hence, in this paper, we use the DaReCzech test set for out-of-domain robustness testing (Section~\ref{sec:dareczech_evaluation}).
DaReCzech development queries with informational intent were allocated to the development set that is used as the stopping criterion during training.
Both the DaReCzech development and test set as well as other parameters are shown in Table~\ref{tab:testset_data}.

\section{Methodology}
\label{sec:methods}

To demonstrate the potential contribution of user behavior data as a complement or replacement of human annotations, we finetune three encoder-only pretrained language models for a relevance ranking task in cross-encoder and bi-encoder settings~\cite{reimers2019sentence}, because both have their specific uses in web search. The inputs consist of a query-document pair, with the document represented by its url, title, and text. We use the aggregated user behavior (clicks and dwell times) along with the document positions from \clickset in order to construct pseudo labels as approximates of human annotations. 

\begin{figure*}
    \centering
    \includegraphics[width=.65\linewidth]{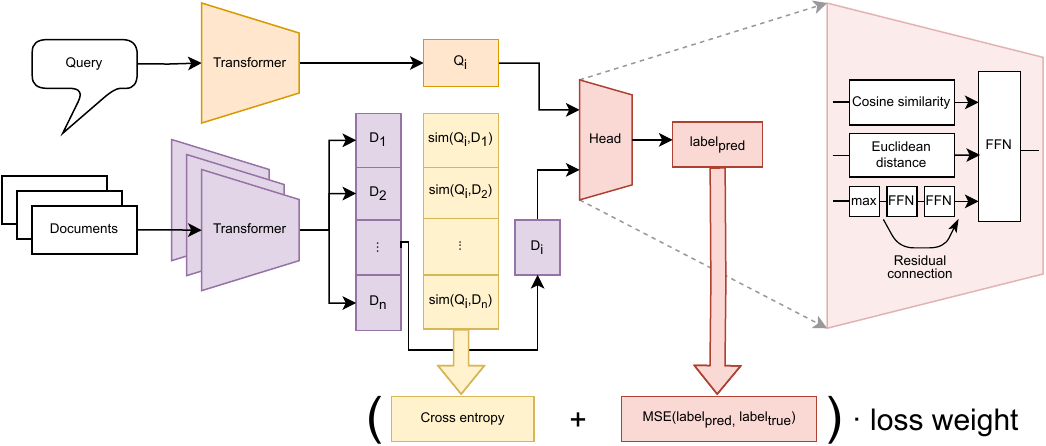}
    \caption{A visualization of the proposed architecture of a bi-encoder relevance model.} 
    \label{fig:architecture-biencoder}
\end{figure*}

\begin{figure}
    \centering
    \includegraphics[width=0.65\linewidth]{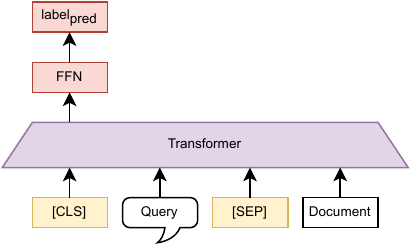}\vspace{-8pt}
    \caption{An illustration of a cross-encoder relevance model.}
    \label{fig:architecture-crossencoder}
\end{figure}

\subsection{Architectures}
\label{sec:architectures}
We conduct training for each model using both cross-encoder and bi-encoder~\cite{reimers2019sentence} configurations. Cross-encoders generally yield superior outcomes in web ranking tasks; however, their slower processing speed renders them impractical for production environments, where rapid ranking of thousands or millions of query-document pairs is essential. Bi-encoders circumvent this performance bottleneck by computing the embeddings independently for queries and documents.
Nonetheless, cross-encoders retain their significance in the training phase, acting as effective teachers during the training process of bi-encoders.

\subsubsection{Cross-encoder}
The term cross-encoder~\cite{reimers2019sentence} describes an architecture that follows the original approach for sequence classification, e.g., BERT~\cite{devlin-etal-2019-bert}. As illustrated in Figure~\ref{fig:architecture-crossencoder}, the input to this model is a single sequence -- a query and a document separated by the special \texttt{[SEP]} token. To predict the relevance of a query to a document, we add an additional linear layer (FFN) on top of the classification token (\texttt{[CLS]}) embedding with a sigmoid activation ($\mathit{label}_\mathit{pred}$) to project a score between 0 and 1.

\subsubsection{Bi-encoder}
The bi-encoder siamese architecture~\cite{reimers2019sentence} (illustrated in Figure~\ref{fig:architecture-biencoder}, together with the loss computation described later in this section) utilizes an identical pre-trained model to compute the embeddings of the query and the document separately. The embeddings are then processed by the interaction module (``Head'' in Figure~\ref{fig:architecture-biencoder}) introduced by~\citet{SZN_SmallECzech}, which first computes the element-wise maximum of the embeddings and passes it through a 2-layer feed-forward network with a residual connection, concatenates the output with the Euclidean distances and cosine similarity of the two embeddings, and passes the result through a final linear transformation with a sigmoid activation. %

\subsection{Pretrained Models}
We finetune three encoder-only pretrained Czech models: Small-E-Czech~\cite{SZN_SmallECzech}, RetroMAE-Small~\cite{SZN_SomeLikeItSmall}, and Fernet-C5~\cite{fernet}. The model parameters are presented in Table~\ref{tab:models}. 

Of these models, only RetroMAE-Small has been pretrained to produce high-quality sentence embeddings~\cite{SZN_SomeLikeItSmall}, making it particularly well-suited for the bi-encoder architecture. Nonetheless, we demonstrate that the other models without a sentence-embedding pretraining objective can attain competitive performance in the bi-encoder settings.

\begin{table}[]
    \centering
        \caption{Model sizes and corresponding parameters. For each model, the total number of parameters is given, and in the brackets is the number of parameters in the embedding layer and the rest of the parameters.}
        \vspace{-5pt}
    \begin{tabular}{lcc}
        \toprule
         \textbf{Model}  &\textbf{Size [type]}&  \textbf{Number of params.} \\\midrule
         Small-E-Czech~\cite{SZN_SmallECzech} &Small&  \hphantom{0}13M (\hphantom{0}4M+\hphantom{0}9M)   \\
         RetroMAE-Small~\cite{SZN_SomeLikeItSmall}  &Small&  \hphantom{0}24M (\hphantom{0}9M+15M)   \\
 FERNET-C5~\cite{fernet} &Base& 162M (85M+77M) \\
        \bottomrule
    \end{tabular}

    \label{tab:models}
\end{table}

\subsection{Training Details}

We train the models utilizing the Adam optimizer~\cite{Kingma2014} using the default hyperparameters and no weight decay. We use accumulated batch size of 500, and learning rates 5e-5 and 2e-5 for small and base models, respectively. For \clickset, we train the models for 4 epochs. For DaReCzech, we increase the number of epochs to 10. With this setup and 4 A100 GPUs, we are able to train a single small cross-encoder model in approximately 1 day, and a single base cross-encoder model in approximately 4 days. The training time is doubled for bi-encoders. Finally, we select the epoch with the highest NDCG@10 performance on DaReCzech dev set queries restricted to informational intent (Section~\ref{sec:dareczech_test_set}).

\subsection{Label Design}
\label{sec:label}

\clickset contains various user behavior information, however, the way of constructing a label as a target model variable for training from this information is not readily apparent. In this Section, we show how clicks, dwell times, and document positions can be used as a relevance label proxy. Ultimately, we propose the resulting most effective formula to combine these three attributes into a single pseudo relevance label, motivated by the correlation analysis carried out in Table \ref{tab:annot_corr}.

\subsubsection{Aggregation}

Many query-document pairs occur multiple times in \clickset. Therefore, when constructing a label for a single query-document pair, we need to aggregate all information available for this pair. To that end, we sum all clicks, dwell times, and ranks for every unique query-document pair, and consider only these sums in the rest of the section. The use of sum over mean or median might seem counter-intuitive, but based on our experiments, sum outperformed other aggregation methods.

\subsubsection{Correlations}

We start by computing the Spearman rank correlation of aggregated clicks, dwell times, click ranks, and their combination with the reference annotations. The reference annotations are obtained by searching all query-document pairs from \clickset in the DaReCzech train set. 

The results are presented in Table \ref{tab:annot_corr}. We see that all considered values show small positive correlation with the reference annotations.
The lower correlation of the raw dwell times with reference annotations is partly caused by missing dwell times for 89.2\% of the pairs, because zero is used for missing values. When mean dwell times are substituted for the missing values, the correlation surpasses clicks. Finally, the combination \textit{click \& dwell time \& rank} based on the formula in Section \ref{sec:clickdwellorder} shows the best correlation with the annotations.

\begin{table}[]
    \centering
        \caption{Spearman correlation between various aspects of user behavior and manual annotations on the DaReCzech train set. Formulas
        in Sections~\ref{sec:order},~\ref{sec:clickdwellorder} are used for rank and click \& dwell time \& rank, respectively.} %
        \vspace{-5pt}
    \begin{tabular}{l c}
         \toprule
         \textbf{Behavior type} & %
         \textbf{Correlation}\\
         \midrule
         Rank & 0.0755 \\
         Dwell Time (NaN$\rightarrow$0) & 0.0884 \\
         Clicks & 0.1335 \\
         Dwell Time (NaN$\rightarrow$mean) & 0.1419 \\
         Dwell Time (NaN$\rightarrow$mean) $\times$ (Clicks $+$ Rank)   & 0.1463   \\
         \bottomrule
    \end{tabular}
    \label{tab:annot_corr}
\end{table}

\subsubsection{Clicks}
\label{sec:click}

When considering clicks, we distinguish between a \textit{last} click in a request and \textit{nonlast} clicks, because the \textit{last} click may indicate either that the user found the required information or abandoned the search. We therefore define weighted clicks as
$$\mathit{wclicks}(q, d) \leftarrow \alpha\cdot\mathit{nonlast\_clicks}(q, d) + \beta\cdot\mathit{last\_clicks}(q, d),$$
where $\alpha$ and $\beta$ are the weights of nonlast and last clicks, respectively.

To construct labels from weighted clicks, we need to map them to a label in the $[0, 1]$ range. We considered monotone mappings, assuming more clicks signify higher relevance, and a $\log$ transformation with a suitable scale delivered the best performance on DaReCzech information-intent development set. We therefore assign the following label to given weighted clicks:
\begin{equation*}
l(q, d) \leftarrow \Big\vert s\cdot\log\big(1+\textit{wclicks}(q, d)\big)\Big\vert_0^1,
\end{equation*}
where $l(q, d)$ is the generated label for query $q$ and document $d$, $\vert x \vert_0^1 = \max(0, \min(1, x))$ clips the input value to the interval $[0, 1]$, and $s$ is a scale factor we describe in Section~\ref{sec:clickdwellorder}.

\subsubsection{Dwell time}
\label{sec:dwelltime}

Dwell time is known only for 39\% of clicks in our dataset, but where available, it provides additional valuable information about user engagement. Analogously to clicks, we obtain the label by transforming the aggregated total dwell time by the $\log$ function:
\begin{equation*}
l(q, d) \leftarrow \Big\vert s\cdot \log\big(1+ \mathit{dwelltimes}(q, d)\big)\Big\vert_0^1.
\end{equation*}

\subsubsection{Rank}
\label{sec:order}
The effect of the rank of a document on its relevance might be equivocal. On one hand, the search engine aims to generate the most relevant documents on top, implying lower rank should indicate higher relevance; on the other hand, mitigating position bias results in increasing relevance for documents with larger rank~\cite{Chuklin2022ClickModels}. Both the correlations and the trained model performance show that the former effect is stronger (lower rank indicating higher relevance), with the following label calculation delivering best results out of the alternatives we considered:
\begin{equation*}
l(q, d) \leftarrow \frac{\mathit{views}(q, d)}{\textit{ranks}(q, d) + C}.
\end{equation*}
The resulting label is a reciprocal of mean rank, with an additional constant $C$ (experimentally chosen as $C=100$) to boost more frequent documents.

\subsubsection{ClickDwellRank}
\label{sec:clickdwellorder}

We also consider combining clicks, dwell times, and rank into a single label. The empirically most successful formula found in our preliminary experiments is the following:
\begin{equation*}
l(q, d) \leftarrow \bigg\vert s\cdot \log\bigg(1 + \scriptstyle \Big(\mathit{wclicks}(q, d) + \frac{\scriptstyle\mathit{views}(q,d)}{\scriptstyle\mathit{ranks}(q, d)+C}\Big) \cdot \Big\vert\mathit{dwelltimes}(q, d)\Big\vert_1^\infty\bigg)\bigg\vert^1_0.
\end{equation*}
The ranks in this formula work mostly as a tie breaker, in case when two documents receive the same amount of clicks and dwell time. The document with higher position (lower rank) then obtains a higher label.

Finally, we set the scale factor $s$ so that virtually all labels are less than 1 and do not need to be clipped, choosing $s=1/20$. For simplicity, we employ the same scale factor for all configurations.

\subsubsection{Loss Weights}
\label{sec:weights}

Given that our labels are aggregated across all query-document pairs, we lose the information about prevalence of each such pair. We can restore the information by using loss weights -- for each query-document pair, we multiply its loss by its number of occurrences in the dataset:
\begin{equation*}
\mathit{loss}(q, d) \leftarrow \mathit{loss}(q, d) \cdot \log\big(2 + \mathit{views}(q, d)\big).
\end{equation*}

Apart from the number of occurrences, we could also try to mitigate the natural imbalance between clicks and non-clicks, given that clicks account for only 27.6\% of query-document pairs in our dataset. Therefore, we also examine an alternative formula accentuating clicked results:
\begin{equation*}
\mathit{loss}(q, d) \leftarrow \mathit{loss}(q, d) \cdot \log\big(2 + \mathit{clicks}(q, d)\big).
\end{equation*}
The constant $2$ is needed in order for the loss weight to be strictly positive even when there are no clicks.

\subsection{Soft Negative Pairs}
\label{sec:soft_negatives_methods}

Because the documents in \clickset are retrieved by a search engine, they are expected to be highly relevant to a given query. However, for successful training, a model might also benefit from clearly non-relevant documents (with relevance label 0). We call such query-document pairs the \textit{soft} negative examples.

A straightforward approach to generate soft negative pairs is to randomly sample a constant number of documents for every query. This approach is naturally applicable in both the cross-encoder and the bi-encoder setting.

\subsubsection{Contrastive Training}
\label{sec:contrastive_training_methods}

In the bi-encoder setting, we can obtain the soft negatives for a given query more efficiently by considering all documents relevant to other queries in the batch -- the so-called \textit{in-batch} negative examples~\cite{gillick-etal-2019-learning}. Specifically, we follow the standard contrastive framework~\cite{chen_contrast} and use the cross-entropy objective with in-batch negatives as a contrastive loss:
\begin{equation*}
\textit{constrastive loss}(q, d) \leftarrow -\log\frac{e^{\operatorname{sim}(q_i, d_i) / \tau}}{\sum_{j=1}^{N} e^{\operatorname{sim}(q_i, d_j) / \tau}}.
\end{equation*}
Here, \textit{N} is the number of documents in the batch, $\tau$ is a temperature parameter that controls the separation of positive and negative examples (our method uses learnable $\tau$ with an initial value of $0.07$), and $\operatorname{sim}(q,d)$ is the similarity measure between the representations of a query $q$ and a document $d$. We specifically use the cosine similarity, hence $$\operatorname{sim}(q,d) = \frac{q^T d}{\|q\| \cdot\|d\|}.$$

\section{Results}
\label{sec:results}

\looseness-1
Our best model uses the label formula presented in Section~\ref{sec:clickdwellorder} along with generated soft negative query-document pairs (Section~\ref{sec:soft_negatives_methods}) and click-based loss weights (Section~\ref{sec:weights}). Furthermore, our bi-encoders also use the contrastive training objective based on the cross-entropy loss with initial temperature $\tau = 0.07$ as described in Section~\ref{sec:contrastive_training_methods}. We report model performance in three settings: 
\begin{itemize}
    \item DaReCzech: training solely on human-labeled dataset,
    \item \clickset: training solely on our click dataset,
    \item \clickset + DaReCzech: pretraining on clicks and finetuning on human annotations.
\end{itemize}

We use the standard ranking metric NDCG@10~\cite{NDCG} and treat only labels with numeric value above $0.5$ as relevant. For every configuration, we train a model three times with different random initialization, and report the average and standard deviation.

\looseness-1
The main results are presented in \tablabel~\ref{table:main_intent} and for reference, the random and oracle baselines are also supplied in \tablabel~\ref{table:baselines}. For cross-encoders, our method of click training achieves +2 percent points in NDCG@10 over our human-labeled DaReCzech baseline. In case of bi-encoders, this difference becomes even more prominent, showing +4.5 percent point NDCG@10 improvement. These differences are statistically significant with p-value less than 0.001, using the Monte Carlo permutation test with 1M samples and probability of error at most $10^{-6}$~\cite{fay-mc-permutation-test,gandy-sequential-mc-risk}. When further finetuning the \clickset-trained models on DaReCzech, we see further statistically significant improvements in the cross-encoder setting, but nearly no change in the bi-encoder setting. We note that our strong baselines achieve performance comparable to state of the art, as evidenced by \tablabel~\ref{tab:prior_work}.

\begin{table}[]%
    \caption{Baseline NDCG@10 [\%] on the \clickset test set.}
    \vspace{-5pt}
    \centering %
    \begin{tabular}{lcc}
    \toprule
    \textbf{Dataset} & \textbf{Random baseline} & \textbf{Oracle baseline} \\
 \midrule
    \clickset test & 22.50 & 98.69 \\   
    \bottomrule
    \end{tabular}
    \label{table:baselines}
  \vspace{10pt}
    \centering %
    \caption{Evaluation of the Czech models finetuned on the human annotated DaReCzech train set, the \clickset train set, or both. We show an average and a standard deviation of three runs, and $\boldsymbol{A < B}$ indicates the difference between models trained on different train sets (comparing columns) is statistically significant (p less than 0.001), using the Monte Carlo permutation test with 1M samples and probability of error at most $\boldsymbol{10^{-6}}$~\cite{fay-mc-permutation-test,gandy-sequential-mc-risk}.}
    \vspace{-5pt}
    \begin{tabular}{lccc}
    \toprule
    \textbf{Test Set:}& \multicolumn{3}{c}{\clickset  NDCG@10 [\%]}\\\midrule
    \textbf{Train Set:}& DaReCzech& \clickset& \makecell[l]{\clickset\\DaReCzech}\\\midrule

 \textbf{Bi-encoders:}& & & \\

    Small-E-Czech & 62.09 $\pm$ 0.2 & \llap{{$\boldsymbol{<}$\kern3.5pt}}68.01 $\pm$ 0.1 & 68.43 $\pm$ 0.1\\
    RetroMAE-Small & 64.12 $\pm$ 0.2 & \llap{{$\boldsymbol{<}$\kern3.5pt}}68.73 $\pm$ 0.1 & 68.75 $\pm$ 0.2\\
    FERNET-C5 & 63.09 $\pm$ 0.7 & \llap{{$\boldsymbol{<}$\kern3.5pt}}69.88 $\pm$ 0.1 & 69.89 $\pm$ 1.1\\

\midrule
 \textbf{Cross-encoders:}& & & \\

    Small-E-Czech & 66.41 $\pm$ 0.5 & \llap{{$\boldsymbol{<}$\kern3.5pt}}68.12 $\pm$ 0.5 & \llap{{$\boldsymbol{<}$\kern3.5pt}}69.86 $\pm$ 0.8\\
    RetroMAE-Small & 65.20 $\pm$ 0.4 & \llap{{$\boldsymbol{<}$\kern3.5pt}}67.48 $\pm$ 0.7 & \llap{{$\boldsymbol{<}$\kern3.5pt}}68.70 $\pm$ 0.4\\
    FERNET-C5 & 68.06 $\pm$ 1.1 & \llap{{$\boldsymbol{<}$\kern3.5pt}}70.05 $\pm$ 0.6 & \llap{{$\boldsymbol{<}$\kern3.5pt}}72.35 $\pm$ 0.3\\

    \bottomrule
    \end{tabular}
    \label{table:main_intent}
\end{table}

\begin{figure}
    \centering
    \includegraphics[width=.9\linewidth]{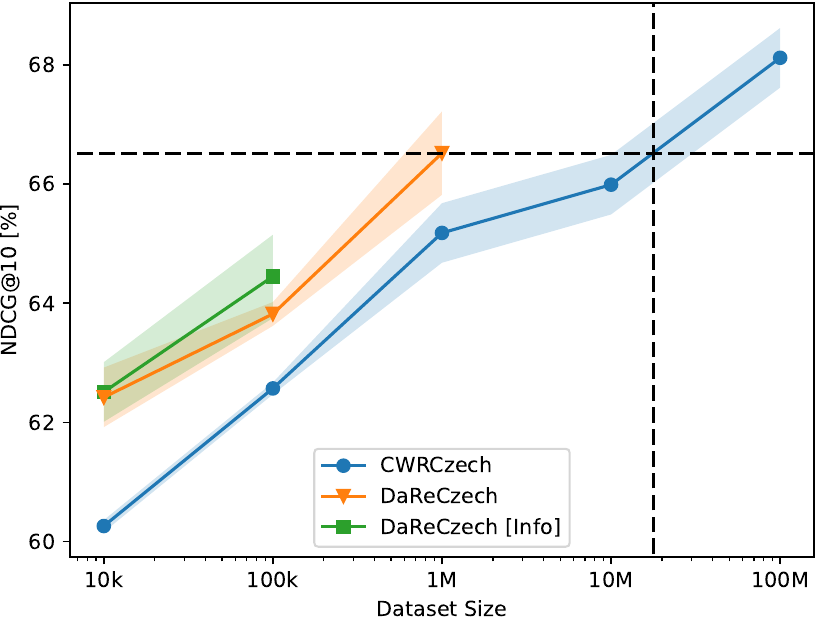}
    \caption{Relationship between user behavior dataset size and Small-E-Czech cross-encoder performance on the \clickset test set. The x-axis represents number of query-document pairs in the dataset used for training.}
    \label{fig:size}
\end{figure}

An interesting question is at what scale the automatically collected data (\clickset) can match the results of human-annotated data (DaReCzech). We show the relation between the amount of click data and performance of our Small-E-Czech cross-encoder model in Figure~\ref{fig:size}. For each dataset size, we sub-sample a random set of queries together with their relevant documents.
The graph clearly shows that in order to match the performance on 1M manual annotations (orange line, DaReCzech), we need around 20M user behavior data (blue line, \clickset), so that both systems trained on their respective datasets reach the same NDCG@10 (circa 66.5; note that the $x$ axis is logarithmic). Furthermore, there appears to be a roughly linear trend, where each order of magnitude increase in training data size results in approximately a 2 percentage points gain in NDCG@10.

\section{Ablation Studies}
\label{sec:ablation}

In this section, we study each component of our final method in isolation.

\subsubsection*{Labels}

In Section \ref{sec:label}, we detailed various methods for label generation, utilizing user clicks, dwell times, document rank, and a combination of these factors. Additionally, for click-based labeling, it is necessary to choose coefficients $(\alpha, \beta)$ for nonlast and last clicks, respectively. For label ablation study, we considered the combinations (1, 1), (0.5, 1), and (1, 0.5), thereby assigning varying levels of importance to the last click relative to the rest. We then train a baseline cross-encoder Small-E-Czech model using each designed label to analyze their respective impacts.

The results are presented in Table~\ref{tab:ablations-cross}.a, ordered by model performance. Our findings indicate that all click- and dwell time-based labels surpass the baseline model which is trained exclusively on position data (row ``Rank''). Notably, the coefficient assigned to the last click plays a significant role, with the best performance surprisingly coming from a model that attenuates its importance (row ``Clicks(1, 0.5)''). Ultimately, the combined label approach (row ``ClickDwellRank'') yields the highest improvement of +1.9 percent points over the position-based label.

\begin{table}[t]
    \centering
    \caption{Ablation experiments of the Small-E-Czech cross-encoder performance measured on the \clickset test set. We show an average and a standard deviation of three runs. (a) The comparison of label design methods (Section~\ref{sec:label}). (b)~The influence of soft negative pairs (Section~\ref{sec:soft_negatives_methods}).}
    \vspace{-5pt}
    \begin{tabular}{lll}
    \toprule
    \textbf{Test Set:}& \multicolumn{2}{c}{\clickset  NDCG@10 [\%]}\\\midrule
    \textbf{Train Set:}& \clickset & \makecell[l]{\clickset\\DaReCzech}\\\midrule
  \textbf{(a) QueryDoc Label:} \\[2pt]
    Rank & 65.34 $\pm$ 0.3 & 66.30 $\pm$ 0.6\\
    Dwell Time (NaN$\rightarrow$20) & 65.58 $\pm$ 0.9 & 68.89 $\pm$ 0.4\\
    Clicks(0.5, 1) & 66.18 $\pm$ 0.2 & 67.12 $\pm$ 0.4\\
    Clicks(1, 1) & 66.30 $\pm$ 0.2 & 67.46 $\pm$ 0.4\\
    Clicks(1, 0.5) & 66.51 $\pm$ 0.5 & 67.93 $\pm$ 0.8\\
    ClickDwellRank & 67.22 $\pm$ 0.5 & 69.30 $\pm$ 0.2\\
  \midrule
  \textbf{(b) Soft Negatives:} \\[2pt]
    Not included & 66.30 $\pm$ 0.2 & 67.46 $\pm$ 0.4\\
    Included & 68.35 $\pm$ 0.2 & 69.39 $\pm$ 0.3\\
    
    \bottomrule
    \end{tabular}
    \label{tab:ablations-cross}
\end{table}

\subsubsection*{Soft Negative Pairs}

We demonstrate how extension of the training dataset with automatically generated soft negatives enhances the cross-encoder Small-E-Czech performance in Table~\ref{tab:ablations-cross}.b. The effect is substantial despite the simplicity of the method, yielding a +2 percent point increase in NDCG@10.

\subsubsection*{Contrastive Training}

The effect of employing the additional contrastive loss (Section~\ref{sec:contrastive_training_methods}) in the bi-encoder setting is quantified in Table \ref{tab:ab-contrastive}.a.
Training with additional contrastive loss yields +6 percent points NDCG@10 increase compared to a non-contrastive baseline. Interestingly, extending the dataset with generated soft negatives as in the cross-encoder settings improves the results further by additional +1 percent points.

\begin{table}[] %
    \centering %
    \caption{Small-E-Czech bi-encoder performance trained with (a) several contrastive objectives (Section~\ref{sec:contrastive_training_methods}) and (b) different loss weights (Section~\ref{sec:weights}).}
    \vspace{-5pt}
    \begin{tabular}{lll}
    \toprule
    \textbf{Test Set:}& \multicolumn{2}{c}{\clickset  NDCG@10 [\%]}\\\midrule
    \textbf{Train Set:}& \clickset & \makecell[l]{\clickset\\DaReCzech}\\\midrule

  \textbf{(a) Contrastive Training:} \\[2pt]
    None & 60.22 $\pm$ 0.8 & 63.72 $\pm$ 0.6\\
    Cross-Entropy w/o Head & 66.17 $\pm$ 0.4 & 66.92 $\pm$ 0.5\\
    Cross-Entropy & 66.62 $\pm$ 0.5 & 66.87 $\pm$ 0.3\\
    Cross-Entropy Soft Negatives & 67.73 $\pm$ 0.6 & 67.70 $\pm$ 1.0\\
    \midrule
  \textbf{(b) Loss Weights:} \\[2pt]
    None & 67.73 $\pm$ 0.6 & 67.70 $\pm$ 1.0\\
    Views & 68.21 $\pm$ 0.2 & 68.45 $\pm$ 0.4\\
    Clicks & 68.05 $\pm$ 0.2 & 68.59 $\pm$ 0.5\\

    \bottomrule
    \end{tabular}
    \label{tab:ab-contrastive}
\end{table}

\begin{table} %
    \centering
    \caption{Out-of-domain performance comparison of models on DaReCzech using the P@10 metric, including prior work in italic.}
    \label{tab:prior_work}
    \vspace{-5pt}
    \begin{tabular}{lcc}
        \toprule
        \textbf{Test Set:} & \multicolumn{2}{c}{DaReCzech P@10 [\%]} \\
        \midrule
        \textbf{Train Set:} & DaReCzech & \makecell[c]{\clickset\\DaReCzech} \\        
        \midrule

  \textbf{Bi-encoders:} \\[2pt]
    \textit{Small-E-Czech}~\cite{SZN_SmallECzech} & \textit{45.26 $\pm$ 0.2} \\
    \textit{RetroMAE-Small}~\cite{SZN_SomeLikeItSmall} & \textit{45.29 $\pm$ 0.3} \\
    \textit{FERNET-C5}~\cite{SZN_SomeLikeItSmall} & \textit{45.87 $\pm$ 0.3} \\
    Small-E-Czech & 45.40 $\pm$ 0.0 & \llap{{$\boldsymbol{<}$\kern3.25pt}}46.19 $\pm$ 0.2\\
    RetroMAE-Small & 45.69 $\pm$ 0.1 & \llap{{$\boldsymbol{<}$\kern3.25pt}}46.38 $\pm$ 0.1\\
    FERNET-C5 & 45.37 $\pm$ 0.2 & \llap{{$\boldsymbol{<}$\kern3.25pt}}46.45 $\pm$ 0.4\\
    \midrule
  \textbf{Cross-encoders:} \\[2pt]
    \textit{Small-E-Czech}~\cite{SZN_SmallECzech} & \textit{46.30 $\pm$ 0.2}\\
    Small-E-Czech & 46.26 $\pm$ 0.1 & 46.43 $\pm$ 0.3\\
    RetroMAE-Small & 46.28 $\pm$ 0.0 & 46.53 $\pm$ 0.2\\
    FERNET-C5 & 46.95 $\pm$ 0.2 & \llap{{$\boldsymbol{<}$\kern3.25pt}}47.40 $\pm$ 0.1\\
    
        \bottomrule
    \end{tabular}
\end{table}

\subsubsection*{Loss Weights}

Since we use aggregated behavior labels, the information about original query-document frequency is lost during training. To restore it, we weight the loss function by the number of views a query-document received as described in Section \ref{sec:weights}. Furthermore, we also consider number of clicks as a loss weight to mitigate the inbalance between clicked and nonclicked documents.%

The results are compared in Table~\ref{tab:ab-contrastive}.b and show that both approaches improve the model performance. Particularly, using views as weights yields the best improvement of +0.5 percent points. %

\subsubsection*{DaReCzech Evaluation}
\label{sec:dareczech_evaluation}

Since our behavioral data were exclusively collected for informational intent, it is reasonable to verify their  robustness and applicability in the out-of-domain setting. For this purpose, we utilize the DaReCzech test set (Section~\ref{sec:dareczech_test_set}) evaluated using the P@10 metric employed in prior work \cite{SZN_SmallECzech,SZN_SomeLikeItSmall}. 
\tablabel~\ref{tab:prior_work} indicates that our models reach the performance comparable to prior work when finetuned on DaReCzech, lending credibility to our primary results. Moreover, it also demonstrates that pretraining on user behavior data enhances performance, even when assessed on data outside the original domain. 

\section{Conclusion}
\label{sec:conclusion}

\looseness-1
We introduced a new click dataset for web relevance ranking in Czech, called \clickset. It features 100M query-document pairs, of which 27.6M are recorded clicks and 10.8M have dwell times, making it a unique resource of this magnitude. Along with the automatically harnessed user behavior data, we also publish a manually annotated test set with nearly 50k query-document pairs. The dataset is available for academic non-commercial use upon request and is subject to license agreement to ensure compliance with GDPR.%

\looseness-1
We also carried out extensive experiments comparing various ways of leveraging user behavior data from the corpus for relevance ranking. Our best model uses a combination of clicks, dwell times, and document rank as a target output variable. It also utilizes generated soft negative query-document pairs for contrastive training, and employs click-based loss weights. This model trained on user behavior data from \clickset achieves 2.5 percent point improvement for cross-encoder training and 4 percent point for bi-encoder training compared to the baseline trained on human annotated data.

Our analysis of the usefulness of the automatically generated data concludes that for Czech relevance ranking, performance on 1M manually annotated data can be matched by roughly 20M of user behavior data and surpassed with higher quantities.

\begin{acks}
This work was partially supported by the Grant Agency of the Czech Republic under the EXPRO program (project No.\ GX20-16819X). 
\end{acks}

\bibliographystyle{ACM-Reference-Format}
\bibliography{references}

\end{document}